# The Distribution of Carbon Monoxide in the Lower Atmosphere of Venus

Daniel V. Cotton[a], Jeremy Bailey[a,d,*], D. Crisp[b,d], V. S. Meadows[c,d]

[a]School of Physics, University of New South Wales, NSW 2052, Australia

* Corresponding Author E-mail address: j.bailey@unsw.edu.au

[b]Jet Propulsion Laboratory, California Institute of Technology, 4800 Oak Grove Drive, Pasadena, California, 91109

[c]Department of Astronomy, University of Washington, Box 351580, Seattle, Washington, 98195

[d]NASA Astrobiology Institute, Virtual Planetary Laboratory Lead Team

Editorial Correspondence to:

Dr Daniel V. Cotton,

c/- Prof Jeremy Bailey,

School of Physics,

University of New South Wales,

NSW 2052,

Australia

Phone:   +61-2-9385-5618

Fax:   +61-2-9385-6060

E-mail: j.bailey@unsw.edu.au

# Abstract

We have obtained spatially resolved near-infrared spectroscopy of the Venus nightside on 15 nights over three observing seasons. We use the depth of the CO absorption band at 2.3 μm to map the two-dimensional distribution of CO across both hemispheres. Radiative transfer models are used to relate the measured CO band depth to the volume mixing ratio of CO. The results confirm previous investigations in showing a general trend of increased CO abundances at around 60 degrees latitude north and south as compared with the equatorial regions. Observations taken over a few nights generally show very similar CO distributions, but significant changes are apparent over longer periods. In past studies it has been assumed that the CO latitudinal variation occurs near 35 km altitude, at which K-band sensitivity to CO is greatest. By modeling the detailed spectrum of the excess CO at high latitudes we show that it occurs at altitudes around 45 km, much higher than has previously been assumed, and that there cannot be significant contribution from levels of 36 km or lower. We suggest that this is most likely due to downwelling of CO-rich gas from the upper atmosphere at these latitudes, with the CO being removed by around 40 km through chemical processes such as the reaction with $SO_3$.

# 1      Introduction

Discovery of the near-infrared "windows" (Allen and Crawford, 1984) allowed information to be gleaned on the lower atmosphere of Venus remotely. The lower atmosphere is otherwise hidden from observation by the vast sulfuric acid cloud deck (extending from ~48 to 70 km altitude) that encircles the planet. Of the available windows, only the K-band 2.3 μm window features a strong absorption from carbon monoxide – in the range 2.295 μm to ~2.4 μm (Allen and Crawford, 1984; Marcq et al., 2005). The 2.3 μm band, when used to infer carbon monoxide (CO) concentration, is most sensitive to a layer of the troposphere just below the sulfuric acid clouds, at 30 to 40 km altitude (Marcq et al., 2005; Tsang et al., 2008).

Carbon monoxide is one of the most important constituents in the complicated chemistry of the Venus atmosphere (Krasnopolsky, 2007). CO is formed in the upper atmosphere by photo-dissociation of $CO_2$. As it descends through the atmosphere, it is lost through chemical processes, which may include a conversion into carbonyl sulfide (OCS) through atmospheric sulfur sinks, and also a conversion back into $CO_2$ (Krasnopolsky, 2007; Tsang et al., 2008; Yung et al., 2009). As CO has a lifetime in the atmosphere of around ten to a hundred days ($10^6$ to $10^7$ s), it is particularly important to studies of atmospheric dynamics (Svedhem et al., 2007; Tsang et al., 2008) in the lower atmosphere.

Measurement of CO abundance below the clouds was first achieved by two probes in 1978. The *Venera-12* mission, before landing at 7 °S, 294 °E, at ~10.30 LT (Moroz, 1981), sampled the atmosphere 8 times between 42 km and ground level, recording ~30 ppmv in the range 42 to 36 km, and ~20 ppmv at 12 km (von Zahn and Moroz, 1985), and being reported as 28 ± 14 ppmv overall (Gelman et al., 1980). Days before, the *Pioneer Venus Large Probe* impacted at 4.4 °N, 304 °E, at 7.38 LT (Moroz, 1981), after sampling CO at three altitudes: 52 km (32.2 ± 7.2 ppmv), 42 km (30.2 ± 2.1 ppmv) and 22 km (19.9 ± 0.4 ppmv) (Oyama et al., 1980). The probe measurements are summarized in Table 1.

*Table 1: Probe based CO mixing ratio measurements with altitude.*

| Reference | Probe | Location, Local Time | Altitude (km) | Mixing Ratio (ppmv) |
|---|---|---|---|---|
| Oyama et al., 1980 | *Pioneer Venus Large Probe* | 4.4 °N, 304 °E, 7.38 LT | 52 | 32.2 ± 7.2 |
| | | | 42 | 30.2 ± 2.1 |
| | | | 22 | 19.9 ± 0.4 |
| von Zahn and Moroz, 1985 | *Venera-12* | 7 °S, 294 °E, ~10.30 LT | 42 to 36 | ~30 |
| | | | 12 | ~20 |

Ground-based observations using the 2.3 µm window, in combination with spectral modeling – using the probe measurements as a basis – first retrieved a tropospheric CO mixing ratio as a linearly interpolated profile: {≤ 22 km: 30 ppmv, 42 km: 45 ppmv, 64 km: 75 ppmv} (~40 ppmv at 35 km) for a 5 arc-second diameter circular section of the nightside disk centered around 30 °N (Bézard et al., 1990). Early maps of the nightside disk revealed no spatial contrast in CO (Crisp et al., 1991). These spectra (Bézard et al., 1990; Crisp et al., 1991) were later reanalyzed in combination to retrieve a linearly interpolated profile (similar to that of Bézard), and reported as a mixing ratio of 23 ± 5 ppmv with a gradient of 1.20 ± 0.45 ppmv/km at 36 km (Pollack et al., 1993).

Variation in CO abundances with latitude was first inferred through analysis of Near-Infrared Mapping Spectrometer (NIMS) data from the *Galileo* flyby of the planet in 1990. North of 47 °N CO abundance was found to be 135 ± 15 % of equatorial values (Collard et al., 1993). Later, the latitudinal variation was confirmed from ground-based observations, as CO mixing ratios across four nightside north-south chords (from two separate observing runs) showed an ~15 % increase between the equatorial regions and latitudes 40 °N/S (Marcq et al., 2005). Later reanalysis of the two chords closest to the limb (from the later observing run) retrieved 24 ± 2 ppmv, at 36 km (with a gradient of 0.6 ± 0.3 ppmv/km), with an increase of 12 ± 4% by 40 °N/S (Marcq et al., 2006). Although the positioning of the four chords from dawn to the limb offered the potential for longitudinal or zonal trends to be revealed, their temporal separation and the uncertainty associated with these measurements prevented this.

Since the arrival of *Venus Express* at the planet in April of 2006, both the VIRTIS-H (Marcq et al., 2008) and VIRTIS-M (Tsang et al. 2008, 2009) instruments have been employed to retrieve tropospheric CO mixing ratios, all assuming linearly interpolated profiles similar to that of Bézard et al. (1990), and all reported as mixing ratios at 35 or 36 km altitude. Sampling at various latitudes with VIRTIS-H revealed that CO abundances increased from the equatorial latitudes to at least 60 °N/S (31 ± 2 ppmv (South), ~40 ppmv (North) as compared to the equatorial 24 ± 3 ppmv) (Marcq et al., 2008). At the time it was noted that dispersion in the measurements suggested a zonal or longitudinal variation of up to 20 % of the latitudinal mean (Marcq et al., 2008). It was also suggested that an asymmetry about the equator, where the minimum actually occurred at 20 °S, could be related to perturbations caused by the high-elevation land mass Aphrodite Terra (Marcq et al., 2008). Interestingly the mirror reverse of this trend, where the minimum is at 18 °N is apparent in the earlier ground-based data but not discussed (Marcq et al. 2005, 2006); fewer significant mountainous regions are present at northern equatorial latitudes.

VIRTIS-M data was used to produce two near-full (and two partial) maps of the southern hemisphere nightside. These maps revealed that not only was CO increasing in abundance from the equator towards the poles, but that it reached a peak near 60 °S (23 ± 2 ppmv at the equator to 32 ± 2 ppmv in the polar collar), and decreased at higher latitudes. An enhancement from dawn to dusk was also seen in the two near-full maps (Tsang et al., 2008). The overturning of the CO past 60 °S was explained as corresponding to the downward branch of the Hadley cell, while the dawn to dusk trend was initially ascribed to a combination of photo-chemistry and dynamical processes (Tsang et al., 2008). However, the most recent work, which includes North-South trace and southern hemisphere map data, has revealed temporal/zonal variability – and on a fast temporal scale not previously seen. This appears to contradict the interpretation of the dawn to dusk trend, and introduces a possible North-South asymmetry (Tsang et al., 2009). The North-South asymmetry may not be real, but an artifact resulting from a combination of relatively sparse northern hemisphere data and the temporal/zonal variation (Tsang et al., 2009). The nature of the temporal/zonal variation is not well understood, but may be due to a complicated interplay of chemistry, atmospheric dynamics and surface topography (Tsang et al., 2009).

*Table 2: A summary of key quantitative determinations of CO mixing ratio at 35 to 36 km using K-band observations and radiative transfer modelling prior to Tsang et al., 2009.*

| Reference | Latitude | Mixing Ratio at 35-36 km (ppmv) | Altitude Gradient (ppmv/km) |
|---|---|---|---|
| Bézard et al., 1990 | 30 ºN | ~40 | |
| Pollack et al., 1993 | Whole Disk, bias N | 23 ± 5 | 1.20 ± 0.45 |
| Marcq et al., 2006 | Equatorial<br>40 ºN/S | 24 ± 2<br>27 ± 3 | 0.6 ± 0.3 (overall) |
| Marcq et al., 2008 | Equatorial<br>60 ºN<br>60 ºS | 24 ± 3<br>~40<br>31 ± 2 | |
| Tsang et al., 2008 | Equatorial<br>60 ºS<br>80 ºS | 23 ± 2<br>32 ± 2<br>~28 | |

Here we build on previous work by presenting for the first time Venus nightside CO maps across the full range of latitudes – extending into the polar regions – providing simultaneous mapping of both hemispheres.

## 2  Observations and Data Reduction

### 2.1  Observations

We used the Anglo-Australian Telescope (AAT) at Siding Spring Observatory with the IRIS2 instrument (Tinney et al., 2004) to acquire spatially resolved spectroscopy of Venus at wavelengths between 2.0 and 2.4 μm in the near-infrared K-band. Details of the instrumental set-up and observing method are described by Bailey et al., (2008a, 2008b). Observing opportunities are limited to month long periods before and after inferior conjunction since this gives the most complete view of the nightside of the planet. We scan the spectrograph slit (7.5 arc minutes long by 1 arc second wide) across the disk, typically ~40 arc sec in apparent diameter, to build up a spectral cube that has a spectral resolution of R ~2400 and two spatial dimensions with a spatial resolution of 0.4486 arc-sec/pixel. This allows us to observe both hemispheres including polar regions and the equatorial- and mid-latitudes simultaneously.

Results from four observing sessions are presented here: July 2004, December 2005, Early July 2007 and Late July 2007. Some scans were excluded due to the deleterious affects of telluric cloud. The remaining scans used for this study are given in Table 3. Typically the scans were acquired in pairs – a scan from East to West across the disk, followed by a West to East scan – known collectively as a *full scan*, though a *half scan* was occasionally taken (East to West), sometimes the affects of cloud cover restricted the useful data to only half of a full scan.

*Table 3: Details of observations.*

| Run | Date (UT) | Scans[a] (No.) | Avg ZD[b] (°) | Apparent Diameter (arc sec) | Apparent Diameter (pixels) | Illuminated Phase (%) | Disk Centre[c] Lat. (°) | Disk Centre[c] Lon. (°E) | Anti-Solar Lon. (°E) |
|---|---|---|---|---|---|---|---|---|---|
| July 2004 | 9-Jul-04 | 3 | 56 | 41 | 91 | 22 | 3.7 | 15 | 91 |
|  | 12-Jul-04 | 2 | 56 | 39 | 87 | 25 | 3.8 | 20 | 98 |
|  | 14-Jul-04 | 4 | 56 | 38 | 84 | 26 | 3.8 | 24 | 102 |
| December 2005 | 8-Dec-05 | 2 | 46 | 40 | 89 | 27 | 2.3 | 286 | 188 |
|  | 9-Dec-05 | 4 | 50 | 41 | 91 | 26 | 2.1 | 288 | 189 |
|  | 10-Dec-05 | 4 | 49 | 41 | 92 | 25 | 2.0 | 290 | 191 |
|  | 11-Dec-05 | 4 | 47 | 42 | 94 | 25 | 1.8 | 292 | 193 |
| Early July 2007 | 1-Jul-07 | 1 | 56 | 31 | 70 | 36 | -0.2 | 255 | 183 |
|  | 2-Jul-07 | 2 | 56 | 32 | 71 | 35 | 0.0 | 257 | 186 |
|  | 3-Jul-07 | 2 | 57 | 32 | 72 | 34 | 0.1 | 259 | 188 |
|  | 4-Jul-07 | 4 | 52 | 33 | 73 | 33 | 0.3 | 261 | 191 |
| Late July 2007 | 21-Jul-07 | 6 | 52 | 43 | 95 | 19 | 3.2 | 295 | 229 |
|  | 22-Jul-07 | 4 | 53 | 43 | 97 | 18 | 3.4 | 296 | 231 |
|  | 26-Jul-07 | 5 | 54 | 46 | 103 | 15 | 4.3 | 303 | 240 |

[a] Refers to the number of scans across the disk (a half scan is one scan, a full scan is two scans).

[b] Angular distance between the observation and the zenith.

[c] To compute local time (in units of longitude), subtract the anti-solar longitude from the longitude.

## 2.2 Data Reduction

First, standard spectral reduction techniques are used to remove the instrumental response. The observations are divided by the flat field image of a quartz lamp. A dark image, obtained with no illumination on the detector, is used to identify hot and dead pixels; which are then replaced by the averages of those surrounding them in the image. A spectrum of a Xenon lamp is used to calibrate the wavelength scale of the spectra.

Sky subtraction and scattered light removal was then performed using techniques based on those described in Meadows (1994). The sky background is removed in a two step process. The primary sky subtraction uses the images at the start and end of each scan, which are assumed to be blank sky frames, subtracting them from each frame in the cube. This approach allows for changing sky levels during the scan (e.g. due to twilight) by making a scaled combination of the two sky frames based on the brightness of a reference sky region near the edge of the frame, away from the planetary disk. This usually leaves a little residual sky, which is removed by a further subtraction based on the spectra near the top and bottom of each frame, again away from the planetary disk. These are extracted, added together and the resulting spectrum subtracted from each row of the frame.

Standard star spectra, as per Table 4, are used to calibrate the spatial slope of the spectrum across the detector.

*Table 4: Standard Star Observations*

| Run | Obs. Date (UT) | Standard Star | Avg. ZD[a] (°) |
|---|---|---|---|
| July 2004 | 13-Jul-04 | BS996 | 45 |
| December 2005 | 11-Dec-05 | BS8477 | 13 |
| Early July 2007 | 1-Jul-07 | BS3421 | 66 |
| Late July 2007 | 26-Jul-07 | BS4013 | 58 |

[a] Angular distance between the observation and the zenith.

Since the dayside crescent of Venus is brilliant compared to the nightside, scattered light from the crescent contaminates the nightside spectra. Scattered light appears as a background solar-like spectrum all across the disk. To remove this contamination we derive a scattered crescent spectrum from the rows just above/below the nightside disk nearest the crescent. We subtract a scaled version of this from each spatial pixel in the cube, with the scaling factor determined at the short-wavelength end of the spectrum, where $CO_2$ absorption in the Venus atmosphere reduces the signal from the Venus nightside to zero, leaving only scattered light (the amount of thermal radiation removed is negligible). This step is challenging and because the region of the nightside directly adjacent to, and contained within the curve of the crescent is strongly affected by scattered light, we consider the data there to be less reliable, and consider only the nightside region from the far nightside limb to the cusps of the crescent when analyzing the data.

By taking the ratio of two wavelength ranges in the K-band, one associated with CO absorption (2.32 to 2.33 μm) and one outside the CO band (2.28 to 2.29 μm) we define a *CO Index* – a relative measure of CO concentration – for each spatial point in the cube. This approach is fundamentally the same as that recently shown by Tsang et al. (2009) to be robust to cloud and temperature variations in establishing CO concentration for all but very great levels of cloud cover. The off-band window by itself (which we denote CLO) is used as a measure of the cloud cover from the signal intensity, which can then be used as a relative measure of the reliability of data for any given part of the planetary disk. The window wavelength ranges are always measured in the reference frame of Venus, correcting for the Doppler shift of Venus relative to the observer.

The approach traditionally taken to remove the affects of the terrestrial atmosphere is to divide the spectrum of the observed object by a spectrum of a standard star, or an unsaturated portion of light from the Venus dayside. However, due to the presence of unresolved high-resolution structure in molecular absorption features, such an approach leads to large systematic errors in spectral regions where both the observed object and the terrestrial atmosphere have strong spectral features (Bailey et al., 2007). Instead, it is necessary to use a forward-modeling approach.

Here, we extend our preliminary analysis (Cotton and Bailey, 2008) by several further steps. Standard atmospheric radiative transfer models (described in Section 3.1) were used to make some corrections to the CO Index. The net effect of these corrections is to both remove effects due to telluric absorption (such as from methane ($CH_4$) which overlaps the CO band), and correct the CO index for any difference in instrument response between the two bands. We label the CO Index from a Standard Star Observation as $C_A$. A synthetic Solar spectrum[1] is Doppler shifted to correspond to the Standard Star observation and then passed through a model terrestrial atmosphere at a zenith angle corresponding to that of the Standard Star measurement, and the CO Index of this is labeled $C_B$. $C_C$ is the CO Index obtained from a Standard Venus atmosphere model (Section 3.1.2). The model Venus atmosphere is then Doppler shifted to correspond with each observation, corrected for the model terrestrial atmosphere transmission (Section 3.1.1) at the appropriate zenith angle, and smoothed to simulate the observation – the CO Index from this is labeled $C_D$.

Each observed CO Index is then multiplied by $C_C/C_D$ to account for the effects of the terrestrial atmosphere, and then divided by $C_A/C_B$ – which is a further correction to account for differences between the modeled terrestrial atmosphere and the actual atmosphere at the time of observing. Thus the CO Index reported corresponds to that modeled, and may therefore be used in conjunction with the atmospheric radiative transfer model to determine mixing ratios. The resulting CO index is the ratio of the radiances in the two bands and can be directly compared with Venus radiative transfer models.

One further step is to 'mask off' data considered unreliable before the hemispherical data are overlayed onto a latitude-longitude/local time grid. In addition to the area inside the crescent, areas where there is significant cloud cover should also be considered less reliable because of a low signal-to-noise ratio. The crescent light subtraction is difficult to achieve with precision, and areas of low signal-to-noise produce larger scattered light subtraction errors. The signal intensity to use as the mask cut-off is somewhat subjective, however plots of the CO Index against the intensity from the off-band $CO_2$ window for small sections of the planetary disk have been used to aid in the determination. Fig. 1 is representative, it shows a spreading of the CO Index recorded and a general trend toward a lower average CO Index with decreasing signal strength. The chosen mask cut-off of 5000 arbitrary detector units (a.u.) is also shown in Fig. 1. Only those data above the cut-off are included in the CO maps presented in Section 4.

---

[1] Kurucz, R. L. Obtained online 12th of January 2009. The solar irradiance by computation. http://kurucz.harvard.edu/papers/irradiance/solarirr.tab

Page 7

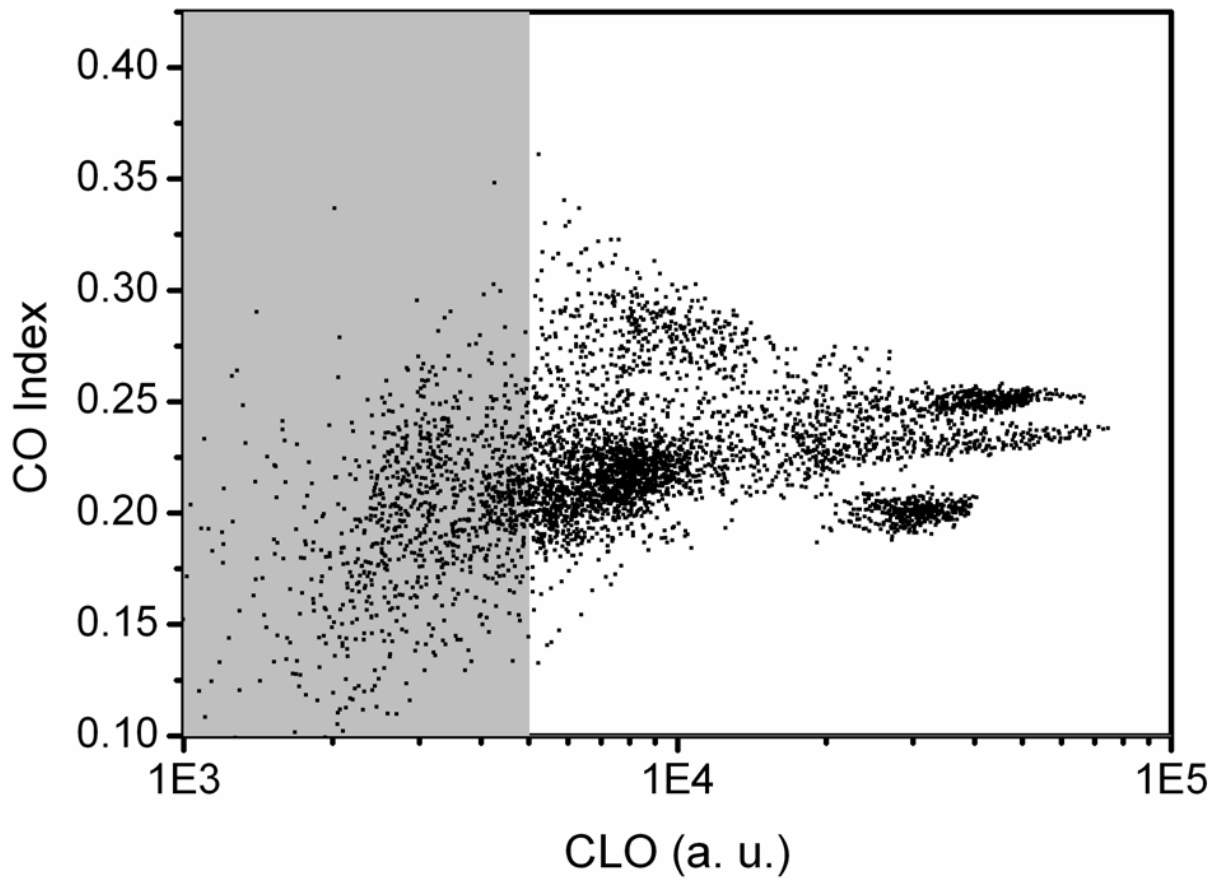

*Fig. 1: CO Index vs. off-band (2.32 to 2.33 μm) intensity (CLO, in arbitrary detector units (a.u.)) for all observations in a 15 x 12 pixel area around the centre of each disk. The grayed out area indicates intensity less than the cut-off applied.*

## 3  Modeling

### 3.1  Standard Models

Here we use the FORTRAN based radiative transfer modeling software VSTAR (Bailey, 2006) to create 1D Venus and Earth atmosphere models.

#### 3.1.1  Earth

Our terrestrial atmosphere model uses a standard atmospheric profile, similar to that described elsewhere (Bailey et al., 2007) tailored to the locality of Siding Spring (a standard VSTAR option). The water vapor content of the standard profile has been adjusted up to a maximum height of 15 km to match the observed $H_2O$ absorption in our standard star spectra. Molecular absorption line data are taken from the HITRAN 2004 database (Rothman et al., 2005). The model includes absorption from $CO_2$, $H_2O$, $O_2$, $CH_4$, CO, $O_3$ and $N_2O$; and Rayleigh scattering in air. The wavenumber grid is 0.02 $cm^{-1}$ in the range 4000 $cm^{-1}$ to 5000 $cm^{-1}$. Model CO Indexes calculated with a wavenumber grid or 0.002 $cm^{-1}$ were not found to be significantly different.

#### 3.1.2  Venus

Our standard model is based on that of Meadows and Crisp (1996). It uses atmospheric pressure and temperature profiles obtained from the Venus International Reference Atmosphere (Seiff et al., 1985), and a

Page 8

Lambertian surface albedo of 0.15 (although the surface is not detected at the wavelengths of these observations). A comparison of $H_2O$ line lists for Venus analysis by Bailey (2009) concluded that BT2 was the most complete and accurate list at Venus lower atmosphere temperatures. Consequently our $H_2O$ absorption line data are taken from the BT2 database (Barber et al., 2006) with an intensity cut-off of $1 \times 10^{-27}$ cm/molecule applied at 700 K. HDO is included at a D/H ratio 120 times the terrestrial. HDO – which is not included in BT2 – and $CO_2$ absorption line data are from the high temperature database described by Pollack et al. (1993). The model uses the Perrin-Hartman sub-Lorentzian line-shape for the far wings of $CO_2$ (Perrin and Hartmann, 1989) with a constant $\chi$ factor at large distances from the line centre (> 300 cm$^{-1}$) as recommended by Meadows and Crisp (1996). These far line wings provide the $CO_2$ continuum absorption, with no additional $CO_2$ continuum being added. The HITRAN 2004 database (Rothman et al., 2005) is used for absorption lines from $N_2O$, CO, $SO_2$, HF, HCl, OCS and $H_2S$.

The sulfuric acid clouds are made up of particles that fall into one of four size regimes – known as modes; from smallest to largest they are m1 (on average 0.6 μm diameter), m2 (2 μm), m2p (3 μm) and m3 (7 μm) (Carlson et al., 1993). Here we use a standard distribution of the four particle modes (Crisp, 1986), encompassing optical density and scattering properties.

Like the model terrestrial atmosphere, we use a wavenumber grid of 0.02 cm$^{-1}$ in the range 4000 cm$^{-1}$ to 5000 cm$^{-1}$. Again, a wavenumber grid of 0.002 cm$^{-1}$ was also trialed, but was not found to result in any significant change to CO Index. A comparison of a typical observational spectrum extracted from a data cube with a VSTAR produced model spectrum is shown in Fig. 2.

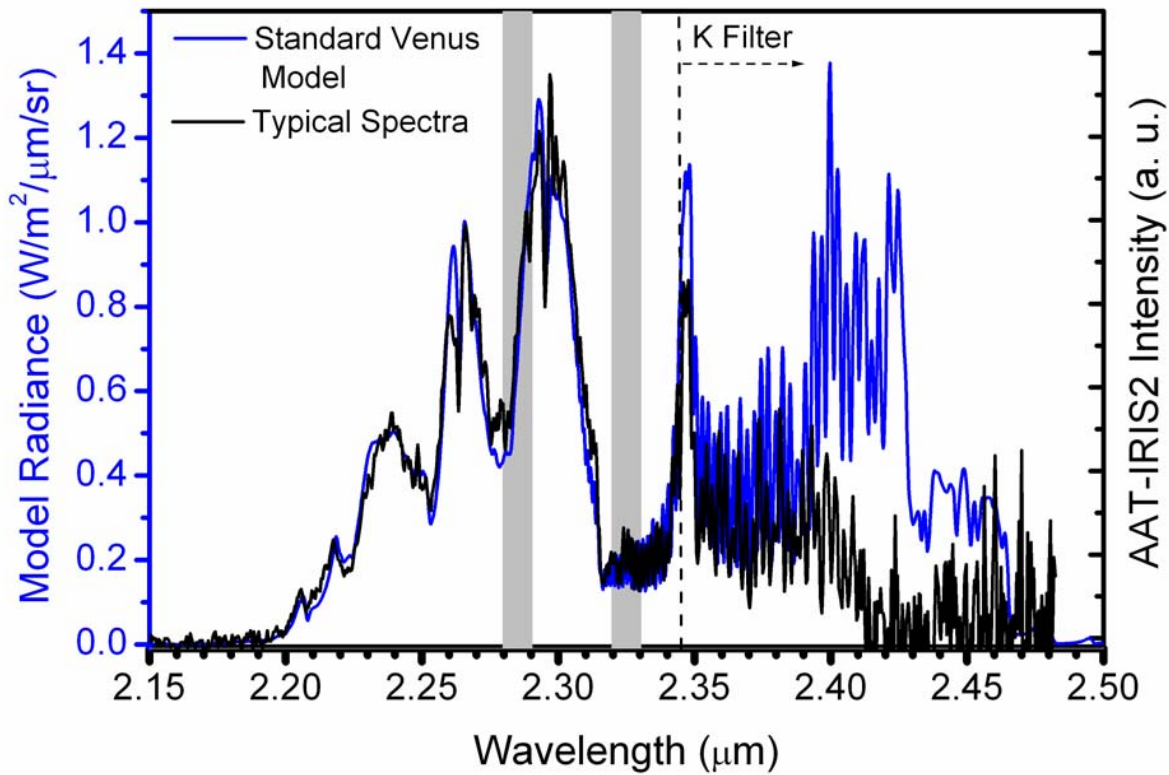

*Fig. 2: A comparison of the standard model spectrum with a typical observation spectrum (southern mid-latitude of a July 26 scan). The dashed line indicates where the K-filter begins to attenuate signal intensity. The grayed out windows are those used for calculating the CO Index and the CLO intensity.*

## 3.2 Conversion of CO Index to CO Mixing Ratio

Here we are primarily interested in the CO mixing ratio, yet what we measure with the telescope is the CO Index (as defined in Section 2.2). To convert between these two quantities we alter the CO mixing ratio in our standard model and calculate the CO Index from the simulated spectra. Like the other minor constituents, the altitude dependence of the nominal CO mixing ratio is that of Meadows and Crisp (1996); it is based on the results of Pollack et al (1993), and at low altitudes has its origin with *Venera* and *Pioneer* probe measurements (Bézard et al., 1990; Kilore et al., 1994), and is therefore most representative of the equatorial latitudes. In this paper, except where explicitly stated otherwise, we have maintained the relative vertical (altitude) profile of the mixing ratio and multiplied it by a consistent factor; we then describe the CO mixing ratio as a percentage of the nominal ratio as well as a mixing ratio at 35 km altitude. Basically the same approach has been employed with little alteration since the first ground-based CO measurements of Bézard et al. (1990). For reference the nominal vertical profile is displayed in Fig. 3; at 35 km the mixing ratio is 23.5 ppmv. We have chosen to represent the CO abundance as the mixing ratio at 35 km so that the results presented here can be compared directly with other works (Marcq et al., 2005, 2006; Pollack et al., 1993; Tsang et al., 2008, 2009) where it has been common to report the mixing ratio at this altitude, as K-band measurements at wavelengths where CO absorbs are typically most sensitive to this altitude. However, the reporting of CO concentration as that at 35 km, shouldn't be taken to mean that this is where variations are occurring. Indeed, later (in Section 4.3) we find this not to be the case.

Overlayed on Fig. 3 is a representation of the altitudinal sensitivity of our measurements. The CO Index generated where 20 % extra CO has been added only in each successive 2 km altitude band is plotted against the altitude of the added CO. Note that the ordinate axis represents the CO Index for the nominal profile. Our choice of CO wavelength window means our CO Index is most sensitive at 33 km, but that 35 km is closer to the centre of the sensitive altitude range. The CO Index is actually sensitive to a wide range of altitudes, with a variation of ~10 % that of the maximum found in the range 20 to 50 km, with altitudes extending to 15 and 65 km still having an influence on the measurement. In particular, it is worth noting that the measurement is still quite sensitive to variations at the bottom of the cloud deck (~48 km).

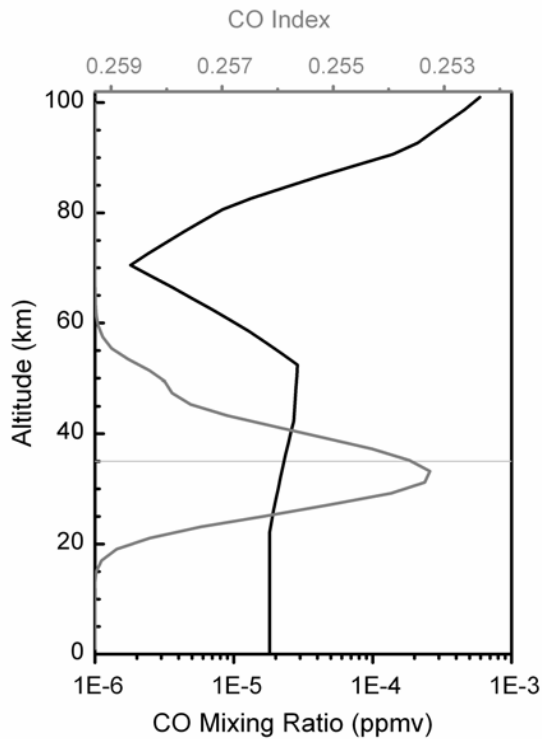

*Fig. 3: The nominal CO mixing ratio (black). Overlayed (gray) is the affect on CO Index of adding 20 % more CO in a 2 km band corresponding to each altitude.*

Fig. 4 illustrates the inverse relationship between CO mixing ratio and the CO Index used for conversion – as the atmospheric CO concentration increases the CO Index decreases in non-linear fashion. An inverse relationship was chosen intentionally to avoid confusion in identifying the scale in subsequent plots.

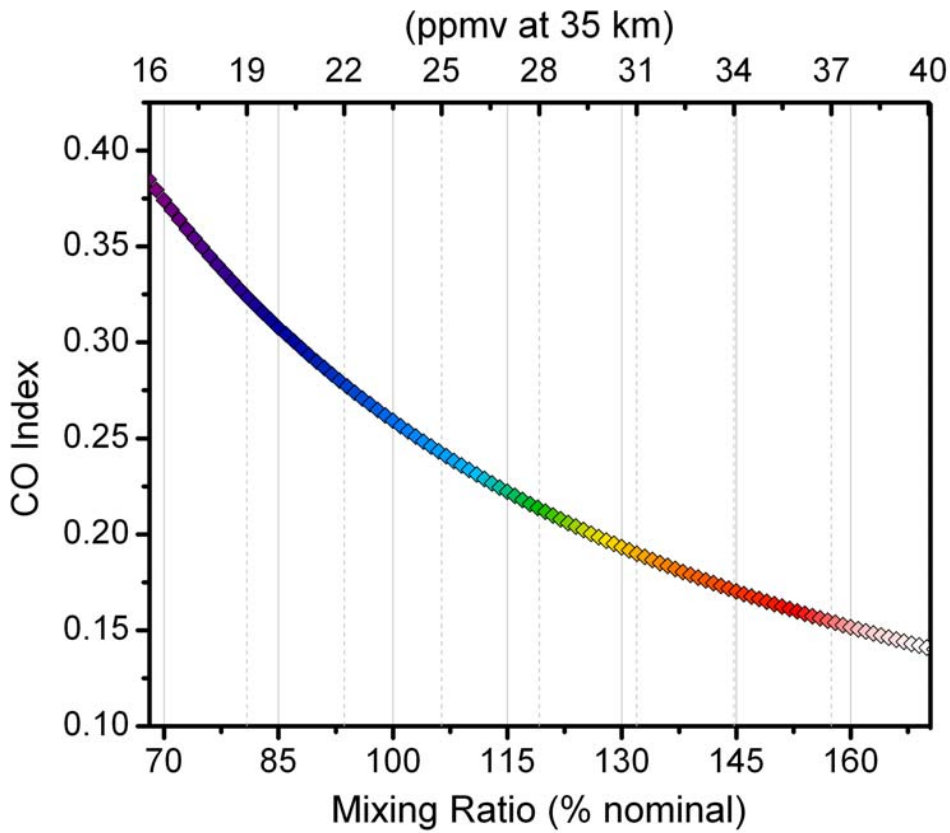

*Fig. 4: How CO Index varies with CO mixing ratio (bottom axis: % of standard ratio, top axis: ppmv by volume at 35 km altitude) The colors are mapped directly to the CO mixing ratio on the abcissa.*

## 3.3 Other Potential Influences on CO Index

Due to limb effects, and changeable cloud cover on Venus, signal intensity varies; the impact this has on the CO Index needs to be quantified. Significant distortions to the relationship between the CO Index and the CO mixing ratio will place limits on the efficacy of the conversion described in Section 3.2, and so need to be investigated.

### 3.3.1 Cloud Cover

Venus' sulfuric acid clouds are well known for their ability to obscure optical observations of the deep atmosphere (Carlson et al., 1993); the variable coverage they provide also affects measurements made using the infrared windows.

Due to Mie scattering the K-band is the IR window most affected by attenuation from the largest particles in the sulfuric acid clouds (Meadows and Crisp, 1996). Tsang et al. (2009) investigated independently the effect of each of the particle modes' opacities on their Radiance Ratio (analogous to our CO Index), finding the difference between them to be of order less than 10 % at useable cloud opacities. Our investigations produced similar results and so are not shown here.

Although Venusian clouds are known to form zonal bands, the bands represent regions where the different mode particles are preponderant rather than exclusive (Carlson et al., 1993). Thus, it is reasonable to expect that any variation in CO Index due to a preponderance of one particle mode over another, in any given region on the disk, is likely to be less than 10 %. Here, for simplicity, we use the standard distribution of the four modes (Crisp, 1986) illustrated in Fig. 5. Though it is important to note that this distribution has a greatest contribution from m3 particles, it gives an opacity intermediate of the extremes of the pure modes, thereby reducing the error in any CO mixing ratio determined from a CO Index affected by the preponderance of a particular particle mode.

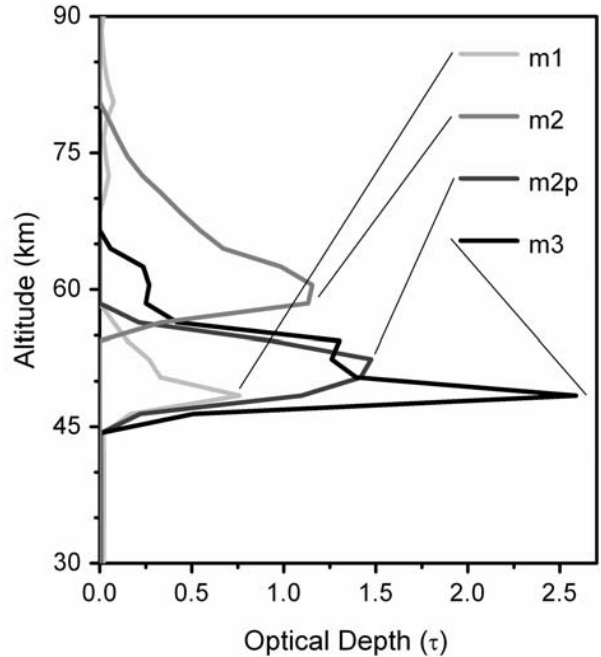

*Fig. 5: The optical depth associated with the different sized sulfuric acid particles in our standard mode distribution.*

We vary the concentration of the cloud particles, fixed in the standard mode distribution, to examine the affects on CO Index and relative signal intensity for a set of different CO mixing ratios in Fig. 5. The curve in these plots – such that they display a minimum in the CO Index – is a result of the wavelength dependant nature of Mie scattering, which is caused by the increasing imaginary component of the refractive index of $H_2SO_4$ with wavelength; the CO window is attenuated slightly more than the shorter wavelength off-band $CO_2$ window, but as absorption begins to saturate in the CO window the trend is reversed. As long as the CLO radiance is above $\sim 4\times 10^{-3}$ W/m$^2$/μm/sr the variation in CO Index results in a variation in determined CO mixing ratio of less than 10 % (or between 2 ppmv and 4 ppmv). While this is the worst case scenario for an individual datum, in practice zero cloud cover is unrealistic and data averaging over a range of radiances reduces the effect. Where the effect is most likely to be of importance is in comparing the July 2004 run with later runs. The July 2004 data is consistently far less attenuated by cloud at equatorial- and mid-latitudes, such that for the CO abundances retrieved, the difference realistically accounts for up to 3 % (~0.75 ppmv) of the variation between this and the other runs (less than the mean difference in CO abundance seen).

Our choice of 2.28 to 2.29 μm for the off-band $CO_2$ window means the value of $4\times 10^{-3}$ W/m$^2$/μm/sr is an order of magnitude higher (on our CLO scale) than the low intensity cut-off described by Tsang et al. (2009), however, the difficulty we faced in crescent subtraction has already forced us to apply a low-intensity cut-off that is likely to be even more conservative than $4\times 10^{-3}$ W/m$^2$/μm/sr. Indeed, inspection of Fig. 1 doesn't reveal any upward swing in CO Index before the 5000 arbitrary detector unit (a.u.) cut-off. Given this is the case, the window 2.28 to 2.29 μm is a sensible choice, as it offers greater sensitivity than a range nearer the 2.30 μm line (Tsang et al., 2009).

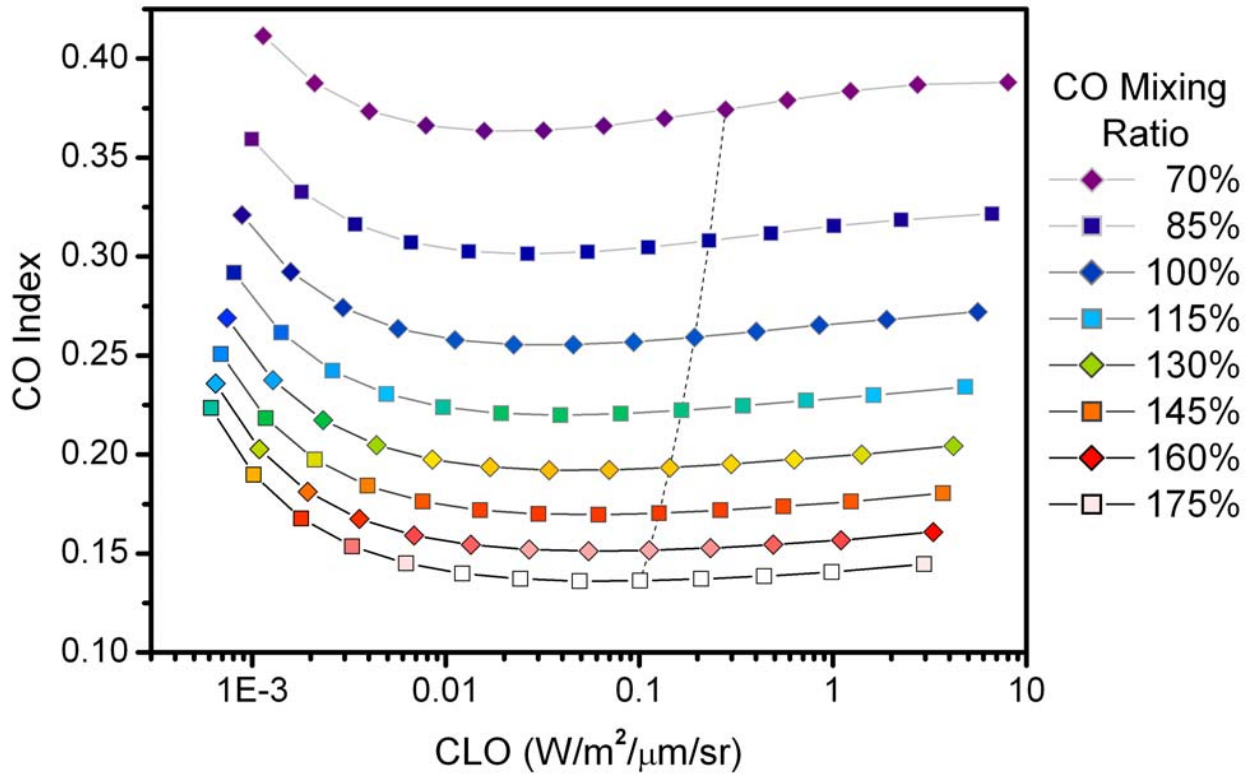

*Fig. 6: Simulated CO Index variation with cloud opacity for a set of different CO mixing ratios. Colors are mapped to CO mixing ratio as per Fig. 4. Each datum in a series is an increment of 25 % from the standard cloud distribution, beginning from 0 % (at right) to 300 % (at left); the dashed line indicates 100 %.*

It is also worth noting that increasing the CO mixing ratio reduces the CLO intensity, since the 2.32 to 2.33 μm window is also attenuated by CO absorption. This can be seen by following the dashed trend line in Fig. 6. Over the range of CO mixing ratios observed – from greatest CO to least – this amounts to an increase in CLO Intensity of a factor of ~3. Consequently, higher CO mixing ratios are slightly more likely to fall below the CLO intensity cut-off we have imposed, and we might expect that, on the average, a marginally lower CO mixing ratio will be retrieved.

### 3.3.2  Limb Effects

As long as a token amount of cloud is included in the model the change in observed CO Index due to emission angle is far less than 1 %. However, as depicted in Fig. 7, a viewing angle associated with the centre of the disk has a CLO radiance ~4 times greater than that at the limb. Consequently, closer to the limb (and therefore toward the poles) fewer measurements meet our low intensity cut-off criteria. If the cloud composition is as we have assumed it to be, then the shape of the curves in Fig. 6 suggests that the CO Index we measure, on average, will be slightly less at the centre of the disk, than for the same mixing ratio observed at the limb. The range of variation is dictated by the relationship between CLO and CO Index (Fig. 6) and thus amounts to less than 10 % error in CO mixing ratio in the worst case scenario, and realistically much less than that.

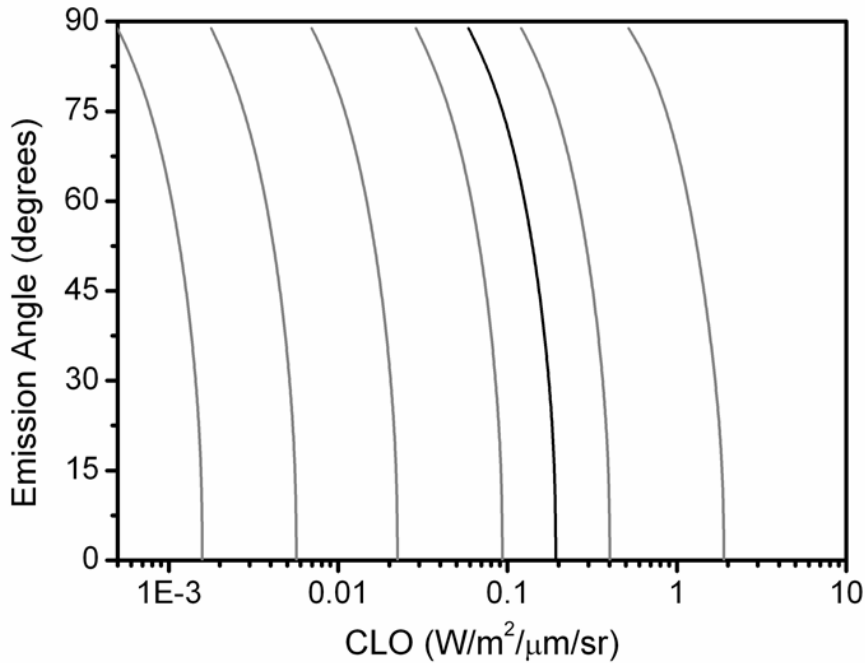

*Fig. 7: CLO intensity with emission angle: the black line represents the model with the standard cloud distribution, while each grey line is an increment of 50 % from the standard cloud distribution, beginning from 25 % (at right) to 275 % (at left).*

### 3.4 Applicability to *Akatsuki*

The Japanese *Akatsuki* orbiter, also known as the *Planet-C*: Venus Climate Orbiter mission, can make use of the method described here with only minor alteration. *Akatsuki* was due to arrive at Venus in December, 2010 (Nakamura et al., 2007), but failed to achieve orbital insertion. However, another attempt may be possible in 2016 (Cyranoski, 2010). The orbiter is equipped with a 2 μm infrared camera (IR2), which has filters centered at 2.26 μm and 2.32 μm, with bandwidths of 0.04 μm and 0.06 μm respectively, intended for CO determination below the clouds (Nakamura et al., 2007).

As can be seen in Fig. 8 the increased separation between the centre of the CO band and the off-band measurement increases the peak-to-trough depth of the curves in the useable range seen in Fig. 6 slightly (a few %) and decreases the peak-to-trough width. Adopting these exact windows for our work would be undesirable, as the maxima and minima would be more likely to correspond to radiance ranges we typically observe greater than the cut-off. Additionally, the CO window includes a small part (2.345 μm – 2.350 μm) that is sensitive to $H_2O$ (Marcq et al., 2008). However, for *Akatsuki* these windows lend themselves to the confident use of a greater radiance range. Thus initial results can be quickly generated from the ratio of the two bands only. If desired, greater precision may be gained by generating the equivalent of Fig. 4 for a range of off-band radiances. Still further improvements may be gained, without resorting to full spectral fitting by employing the 1.735 μm filter to determine the ratio of different mode cloud particles present in the manner of Carson et al. (1993).

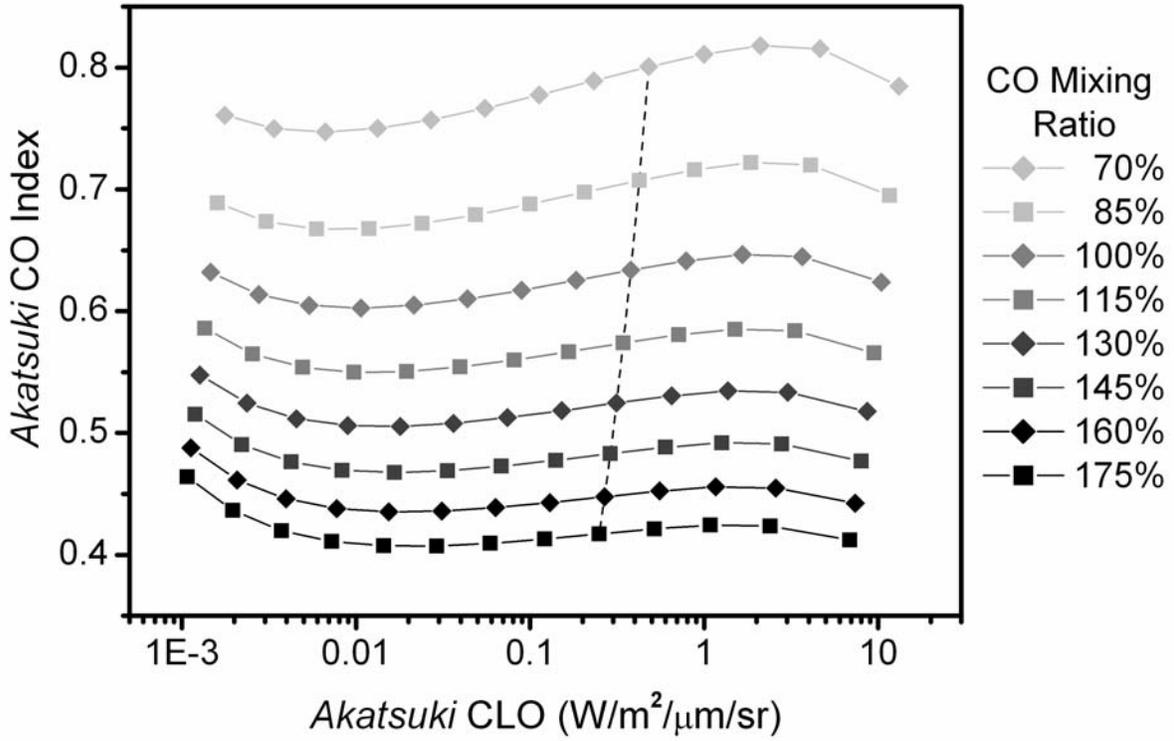

*Fig. 8: Equivalent of Fig. 6 for Akatsuki assuming square windows. Note that no CO ppmv color scale has been applied to this figure.*

# 4 Results
## 4.1 Night Maps

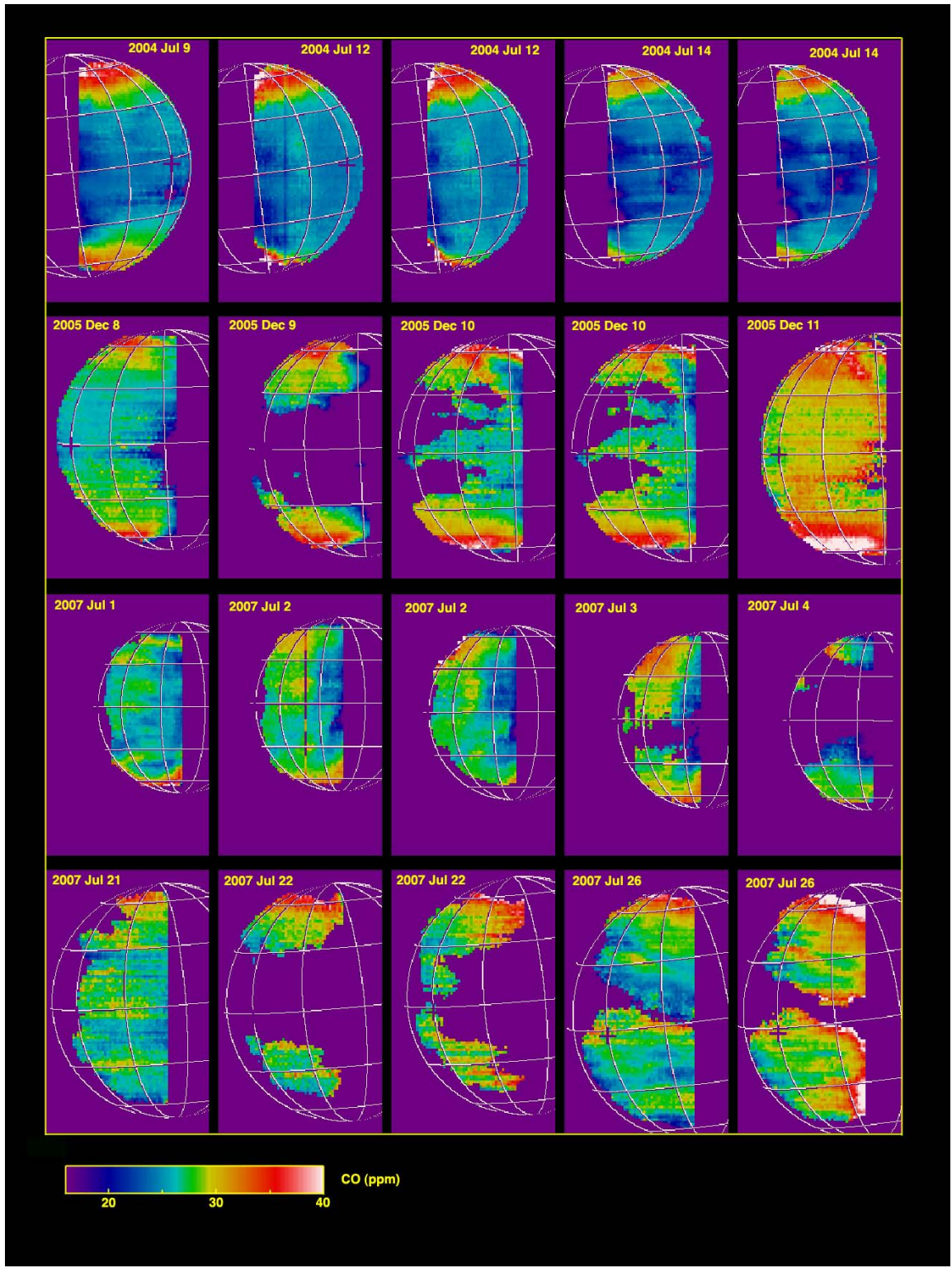

*Fig. 9: Individual maps for each night's observations. Left to right, top to bottom: 2004, Jul 9, Jul 12, Jul 12, Jul 14, Jul 14; 2005, Dec 8, Dec 9, Dec 10, Dec 10, Dec 11; 2007, Jul 1, Jul 2, Jul 2, Jul 3, Jul 4; 2007, Jul 21, Jul 22, Jul 22, Jul 26, Jul 26. Where there is a single map for a given night, that map represents the average of all scans for that evening, where there are two maps the scans have been split to demonstrate data reproducibility.*

Fig. 9 shows the average CO distribution for most days of observation, the exceptions (Jul 12, 2004; Jul 14, 2004; Dec 10, 2005; Jul 2, 2007; Jul 22, 2007; and Jul 26, 2007), have been broken up by "full scans" (see Section 2.1) to show that the data is reproducible, and that changes on the time scale of hours are minor if present. It is possible to see some changes over the course of an observing run – the longest of which is less than a terrestrial week – however, it needs to be remembered that the variability in Venusian cloud, in places, alters the distribution of data we have, and affects minor changes in CO Index measured. Though Dec 11, 2005 shows a similar CO distribution pattern to the others in the December 2005 run, it has an elevated CO mixing ratio across the disk. This may be the result of an earthly contaminant, but is more likely a calibration inconsistency related to the challenging crescent subtraction. With these things in mind, the distribution pattern of CO is quite consistent intra-run.

Fig. 10 shows the CO distribution averaged over each observing run, with distinct differences between the runs, in addition to some common features. The clearest commonality is in the increase in CO toward the polar collars. All four runs show a significant enhancement in CO concentration at the northern polar collar. Additionally, with the exception of the Late July 2007 run (Fig. 10 (d)), which doesn't have data as far south as the others, a clear enhancement at the southern polar collar can also be seen.

The three Venusian midnight to dawn observations (Fig. 10 (b), (c), (d)) also show a drop-off in the CO concentration in one or both polar collars close to dawn. This feature wasn't evident in the two full southern hemisphere maps of Tsang et al. (2008), but is present in one of their partial maps (MTP04-121-0). This may have consequences for Tsang et al.'s (2008) dawn to dusk enhancement, which we discuss in Section 5.5.

Fig. 10 is also striking for the variation between the runs. There is a significant variability between runs in the equatorial- and upper-mid-latitudes. The dawn to midnight run from July 2004 (Fig. 10 (a)), has CO concentrations that are consistent and comparably low all the way up/down to 40 °N/S; a pattern which is a good match for the maps of Tsang et al. (2008), and the chords of Marcq et al. (2006). The three midnight to dusk runs however, show much more variability at equatorial latitudes, and have CO concentrations that begin to increase more sharply toward the collars from 25 or 30 ° – a pattern that has only been indicated previously by the ensemble distribution of Tsang et al. (2009). The map corresponding to December 2005 (Fig. 10 (b)) has a filament of high CO concentration seemingly flowing down from the northern polar collar. Early July 2007, also shows a filament stretching up from the southern polar collar: we cannot find any analogue in the existing CO literature for these.

The most intriguing feature though, is present in the Early July 2004 run (Fig. 10 (d)). Here the CO concentration in the polar collar is severely depleted well before the evening terminator. At the same time we see more than a filament of high CO concentration but rather elevated levels extending all the way from the northern polar collar down to 20 ° at the West of the map, and bleeding through all the way to the equator in the centre of the disk. This unusual pattern is made even more striking by the fact that in the Late July 2004 run, just two terrestrial weeks later, the pattern looks almost completely different. The best evidence for such a pattern having been seen previously in the literature comes from two of the chords from Marcq et al. (2005): their slit position 1 and 2. These two chords have been neglected in later analyses (Marcq et al., 2006; Yung et al., 2009) using the other data from the same paper; nevertheless, they clearly show CO concentrations extending from just north of the equator poleward to the North of ~30 ppmv or more.

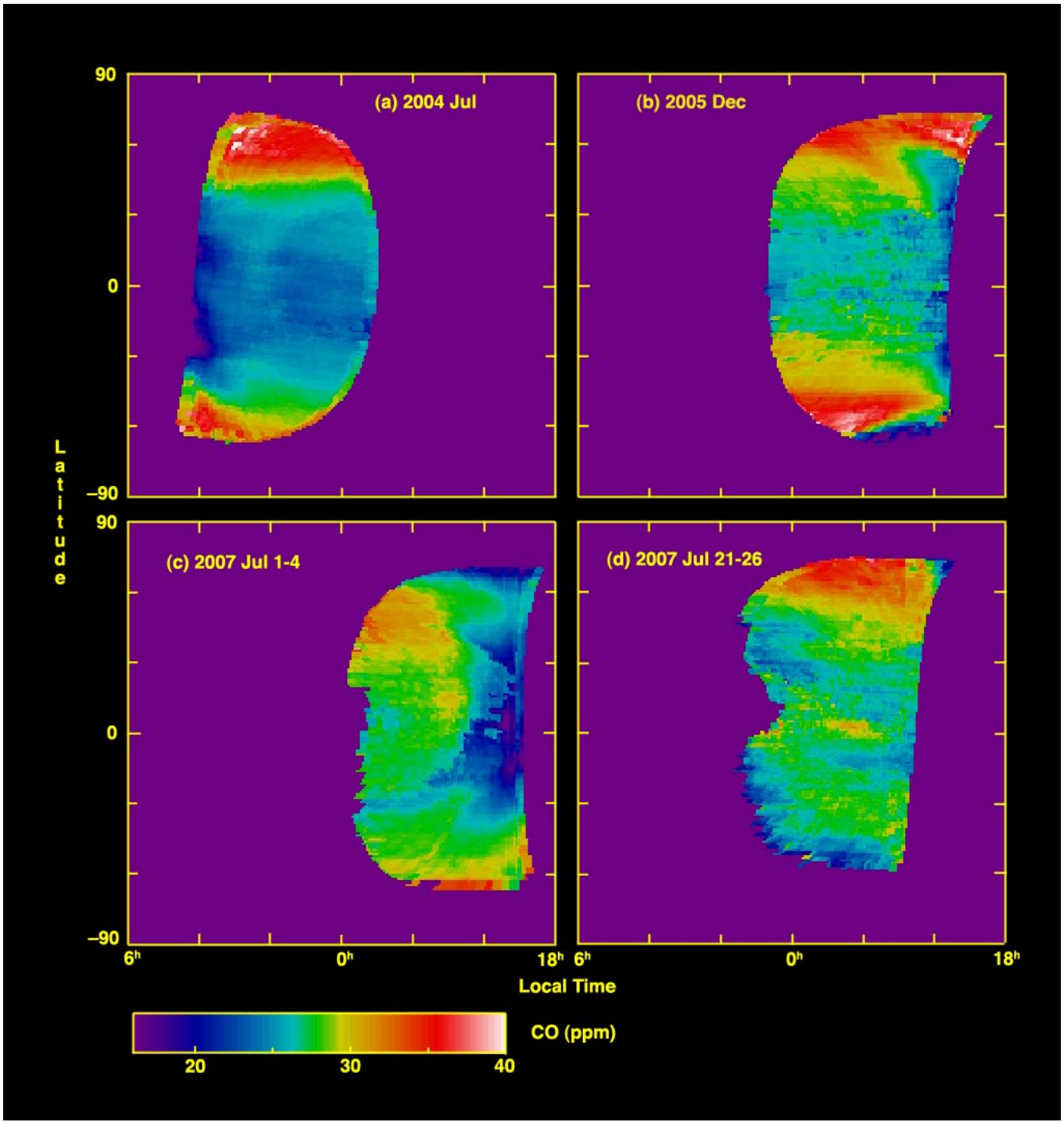

*Fig. 10: CO distribution by observing run: (a) July 2004, (b) December 2005, (c) Early July 2007, (d) Late July 2007. These are produced by projecting the individual CO maps onto a simple cylindrical projection of latitude and local time (thus allowing for the changing apparent size and orientation of Venus) and then merging the individual images by forming the median of all the overlapping maps.*

## 4.2 Latitudinal Variations

Since the most persistent feature in the CO maps was an increase in concentration with latitude toward the polar collars, in Fig. 11 we plot the mean concentration, and ensemble density to examine latitudinal trends with greater precision. An elevated CO mixing ratio from equatorial latitudes toward the polar collars has been observed before for both the North (Collard et al., 1993; Marcq et al., 2005, 2008; Tsang et al., 2009) and the South (Marcq et al., 2005, 2006, 2008; Tsang et al., 2008, 2009); we see the same trend here. We find minima of 24.9 ± 2.4/2.9 ppmv at latitudes 10.5 ºN/S, consistent with past findings, and note a gradual increase toward the polar collars. Because of the relative scarcity of data far beyond 60 º it is hard to pin-point the latitude where, poleward of which, the mixing ratio begins to fall away. However, a judicious examination of Fig. 11 suggests ~56.5 ºS (31.5 ± 4.6 ppmv) and ~61.5 ºN (34.2 ± 5.2 ppmv) are the maxima. These shoulders were first observed using *Venus Express'* VIRTIS-M instrument (Tsang et al., 2008, 2009), and ascribed to the overturning of the atmosphere through the descending branches of the Hadley cells at the polar collars. CO maps (Tsang et al., 2008) produced of the southern hemisphere are particularly definitive in showing the reversal of CO concentration poleward of ~60 ºS.

Also of interest, examining Fig. 11, is the latitude at which CO concentration begins to increase toward the poles. In the South the average concentration is fairly steady until ~25 ºS, whereupon it rises with a consistent gradient to the collar. In the North the situation is not as simple, owing largely to the influence of the Early July 2007 run, the gradient appears to increase steadily from the minimum to ~25 ºN, before increasing again at ~45 ºN.

Due to the orientation of *Venus Express'* orbit there are far fewer observations of the northern hemisphere and hardly any that span a latitudinal range inclusive of both polar collars. Consequently, a suggestion that there exists a North-South asymmetry, with a greater CO concentration in the South, has yet to be verified (Tsang et al., 2009). The data we present here, with much more symmetrical coverage of the two hemispheres, is also North-South antisymmetric, but by contrast, reveals a greater CO concentration in the Northern hemisphere.

Although the 5 ppmv standard deviation in the latitudinal averages at the polar collars might suggest that there is no statistically significant difference between the two hemispheres, we concur with the view of Tsang et al. (2009), that the data spread is largely the result of real zonal variation, and not simply due to measurement uncertainty. This view is supported by Fig. 11, where, even in the densely sampled equatorial latitudes the distribution of data about the mean is non-Gaussian – and in some places even appearing as bi-modal. Additionally, up to 60 ºN/S we have more data poleward than at mid-latitudes (see Fig. 9), yet the clear trend is to greater variance from ~15 ºN/S. Furthermore, one would expect the reverse of this trend if the limb effects described in Section 3.3.2 were significant, or if large variations in cloud cover were playing a role – since the effect on mixing ratio is greater as a percentage for lower concentrations (see Section 3.3.1). Therefore, it is reasonable to conclude that the bulk of variability is due to real zonal or temporal variability.

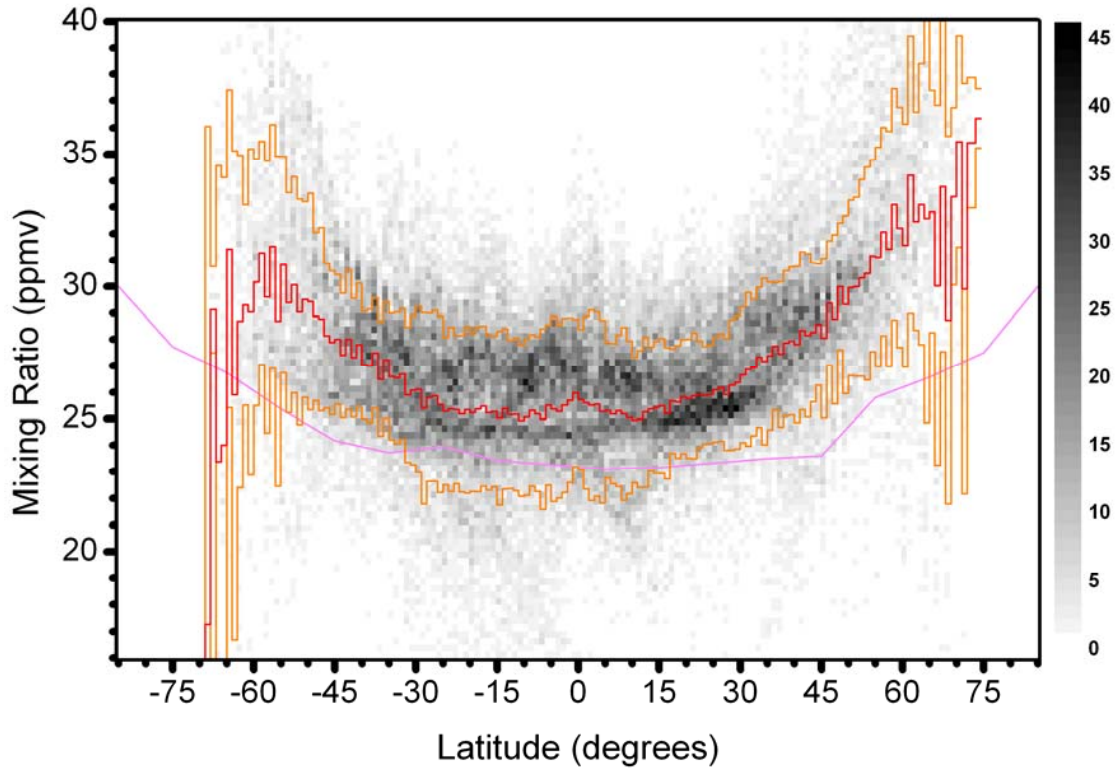

*Fig. 11: Density of data on a grid of 1 degree by 0.25 ppmv. The color scale indicates the number of data points per bin. The colored lines represent the latitudinal averages (red) and ± the standard deviation (orange). The data from all scans presented in Fig. 9 except those from December 11, 2005; July 1, 2007; and 2 of the 5 scans from July 26, 2007 are represented here. Of those excluded, the July 26 scans show unusual effects around the terminator; the July 1 scan shows a distribution not entirely consistent with the others, is the only single half scan and has the lowest spatial resolution; whilst the December 11 scans have an elevated mixing ratios across the whole disk. The pink line is Yung et al.'s (2009) 'Model F' prediction of CO.*

## 4.3    Altitude of the CO Changes

Previous works (Marcq et al., 2005, 2006, 2008; Tsang et al., 2008, 2009) have made the assumption that variations seen in CO concentration occur at 35 km altitude since this is where the measurement is most sensitive. However, as can be seen in Fig. 3, the measurement is actually sensitive to a wide range of altitudes. Spectra with our resolution are quite sensitive to the actual altitude of the CO. In past studies Pollack et al., (1993) and Marcq et al. (2006) have fitted spectroscopic data with both a mixing ratio at 36 km and a vertical gradient in ppmv/km. These results suggest that the CO mixing ratio increases slowly with altitude at $1.20 \pm 0.45$ ppmv/km according to Pollack et al. (1993) and $0.6 \pm 0.3$ ppmv/km according to Marcq et al. (2006).

Marcq et al. (2006) also analysed two traces, one at mid-latitudes (20 °S – 40 °S) and one at equatorial latitudes (0 – 20 °S) separately; they found a greater positive gradient for the mid-latitude position ( ~0.8 ppmv/km compared to ~0.3 ppmv/km). In doing so they note that this information could also be interpreted as a shift in the reference level used for CO mixing ratio determination. This lead to an alternate reference level of 37 km, but relied upon the assumption that the new level was close to the original reference level of 35 km.

However, there have been no in depth investigations of whether the nominal distribution applies to the enhanced CO levels seen at high latitudes. We can use our data to probe this directly. A sensitive differential measurement is possible by taking the spectrum in a high CO (high latitude) region and dividing this by a spectrum in a region with a lower level of CO near the equator. By doing this we remove the effects of instrument sensitivity and telluric absorption, and leave a spectrum, which is that of the additional absorption caused by the extra CO. This can be compared directly with models.

Fig. 12 shows models of the differential effect of adding 20 % extra CO in various different 2 km high layers. The spectra are generated by running a Venus VSTAR model, as described earlier, with the CO mixing ratio increased by 20 % in one layer. The resulting spectrum is convolved with the instrumental point spread function (PSF) to generate a spectrum with the same resolution as the data. This is then divided by a similar spectrum with the nominal CO mixing ratios (it is important to do these steps in the right sequence – dividing the modeled high-resolution spectra and then convolving with the spectrograph PSF produces a quite different result). Note that the effect is complicated by the fact that most of the CO lines are already saturated in the line core with the standard CO profile. This means that the cores of lines from high altitude CO cannot add additional CO absorption, because there is no light to absorb. Only the wings of the lines can contribute additional absorption.

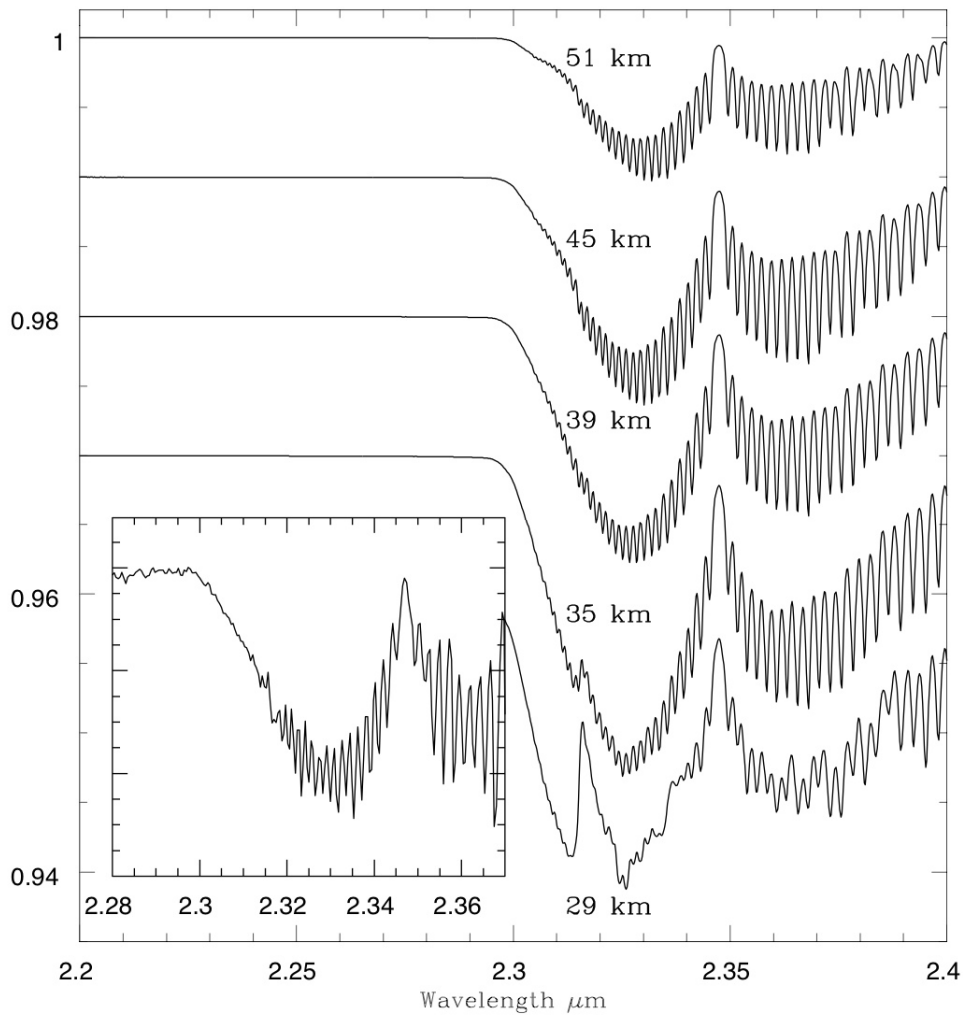

*Fig. 12: The differential effect of adding 20 % extra CO in selected 2 km wide layers centered on the labeled altitude (modeled spectrum with extra CO divided by spectrum without); the ratio of each is successively offset by 0.01 for every 2 km below 51 km for clarity. The 51 km and 45 km spectra are scaled up by a factor of two compared with the others. The inset shows the observed spectrum of a high CO high latitude region (northern polar regions of July 2004 data) divided by a lower CO equatorial region.*

Page 21

These results show that the spectrum of additional CO absorption is different depending on the altitude of the additional CO. The main effect we are seeing is due to the increasing temperature at lower altitudes. As temperature increases more absorption is shifted to higher J levels in the sequence of rotational lines. This changes the overall shape of the CO band pushing the peak further from the 2.345 μm band centre, and in particular, it increases the steepness of the edge of the band at about 2.3 to 2.31 μm (in contrast an increase in the CO gradient as per Marcq et al. (2006) increases the steepness at the band edge only after 2.31 μm). At lower altitudes further effects start to contribute. Absorptions of other species start to become significant, and in particular there is a $CO_2$ absorption feature at about 2.316 μm that reduces our sensitivity to CO at low altitudes. In addition increased pressure broadening of the CO lines and the presence of additional lines from hot CO bands has the effect of reducing the amplitude of the CO line features.

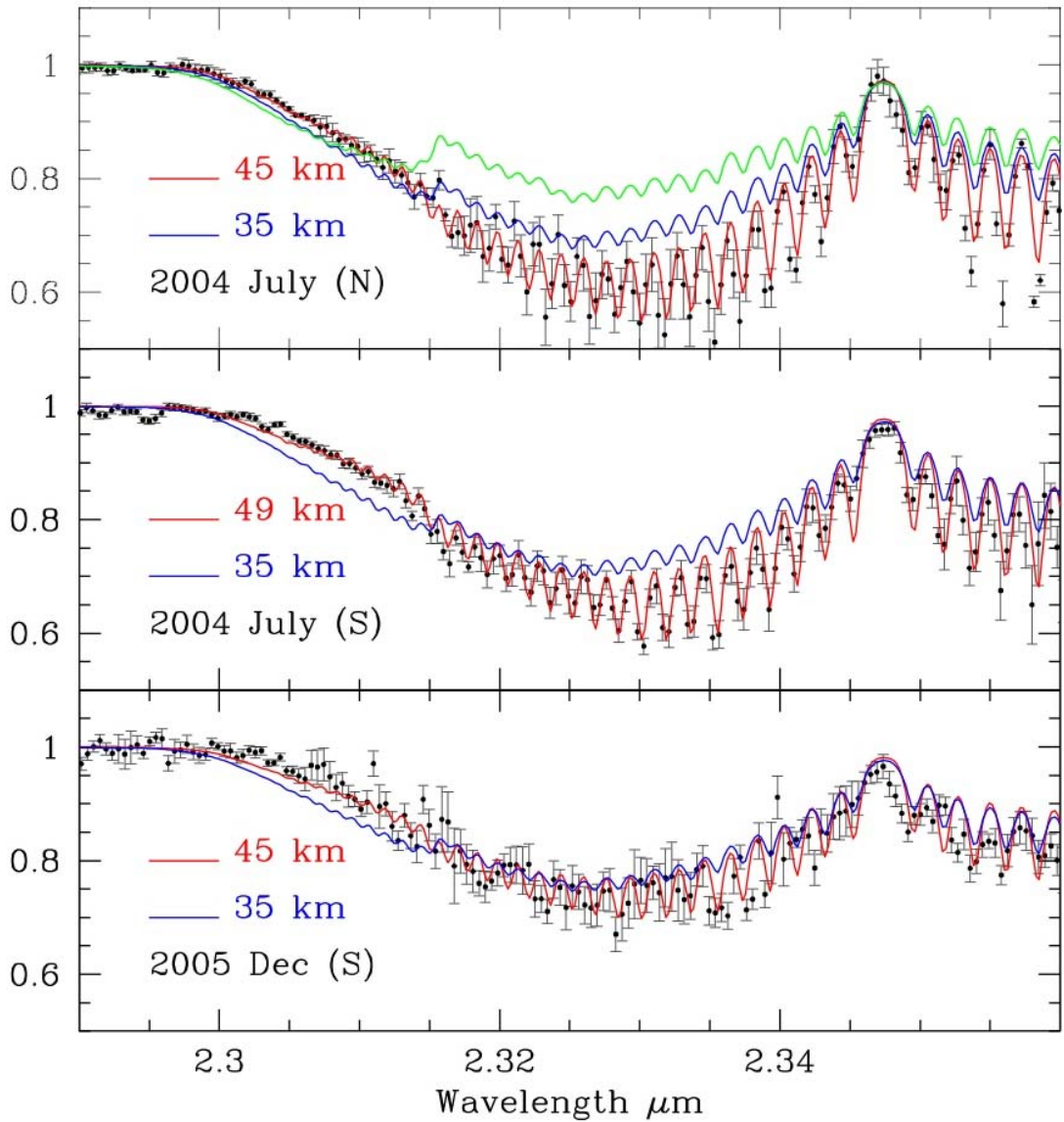

*Fig. 13: Observed spectra of high CO regions divided by low CO regions (dotted line) compared with modeled ratio spectra (scaled to fit the observations) as in Fig. 12. The blue line is a model with CO added to the 35 km layer. The red line is the best fitting model which has the CO at 45 km for the top and bottom panels and at 49 km for the middle panel. The green line (top panel only) is a model in which the CO is increased by the same factor at each altitude.*

To derive an equivalent spectrum from the data we take the observed spectrum in a region of high CO index (generally at high latitudes) and divide by the observed spectrum from a low CO region near the equator. The best dataset for this purpose is that from July 2004 where the CO variations are substantial at both poles, and the cloud opacity was low allowing the use of data from much of the disk. To improve S/N we combined data from four individual maps over July 9-14 and used large rectangular regions, extracting the spectrum from much of the high-CO regions near the poles, and a comparable region near the equator. The resulting ratio spectrum from the northern high-CO region is shown in the inset of Fig. 12 and in the top panel of Fig. 13.

In Fig. 13 the data points are shown with error bars derived from the statistics of the individual spectra that were combined to form each of the three spectra. The model spectra have been fitted to the observations by scaling to minimize $\chi^2$ over the wavelength range 2.29 to 2.35 µm. It can be seen that the observed spectrum does not match that expected from CO near 35 km altitude (green line). If this spectrum is adjusted to fit the 2.3 to 2.31 µm region (which is what the $\chi^2$ minimization tends to do as the errors are smaller here) it has too little absorption over 2.32 to 2.34 µm. The amplitude of the CO line sequence is too small and the model poorly matches the overall shape of the data over 2.3 to 2.35 µm. The reduced $\chi^2$ of this model is 6.65. A model in which CO is increased by the same percentage at all altitudes (green line in Fig. 13. top panel) is an even worse fit with a reduced $\chi^2$ of 16.6. It has a steeper edge and a strong 2.316 µm feature that are not seen in the data. An excellent fit to the data, is however, obtained with CO added to the 45 km layer as shown in the top panel of Fig. 13 (red line) giving a reduced $\chi^2$ of 1.51. An analysis of a more extensive set of models adding the CO to each atmospheric layer gives a best fit with CO at 45 km with an error of ±5 km. Models with the excess CO at 35 km or lower are excluded at the 98 % confidence level.

The middle panel of Fig. 13 shows the data for the southern high-CO region of the July 2004 data. This has similar spectral structure but is best fitted by a model with the CO at 49±5 km, although the 45 km model also fits acceptably. Similar ratio spectra obtained from all the other observing sessions have much a poorer signal to noise ratio, as the CO contrasts are less marked and these datasets generally had higher cloud opacities and so it was harder to find a good low-CO region to form the ratio. However, the overall spectral shapes seem consistent with those obtained from the July 2004 set. The lower panel of Fig. 13 shows the ratio spectra from the southern high-CO region of the December 2005 data. While this is quite noisy and the individual CO lines are not clearly resolved, the overall shape and the edge of the band are still better fitted with a model with CO at 45±8 km.

## 5      Discussion

### 5.1    Latitude Distribution and Variability

Our results confirm previous studies in showing a general trend for high CO concentrations to be seen at high latitudes (around 60 ºN/S). However the details of this pattern show considerable variability. The Late July 2007 data show no increase in the South, although this dataset does not extend quite as far south as some of the others. The Early July 2007 data shows a patch of high CO concentration at lower latitudes, and both this and Dec 2005 dataset show filaments of high CO extending southwards. Where we have observations over several nights the distributions of CO are found to be quite similar. In particular we do not see any significant rotation of the features that are well defined in longitude, showing that the CO does not participate in the atmospheric super-rotation that is seen at the cloud level.

While the distribution does not change much over a few days, the pattern is distinctly different in the observations from Early July and Late July 2007. Hence variability in the distribution is occurring on a time scale of ~20 days.

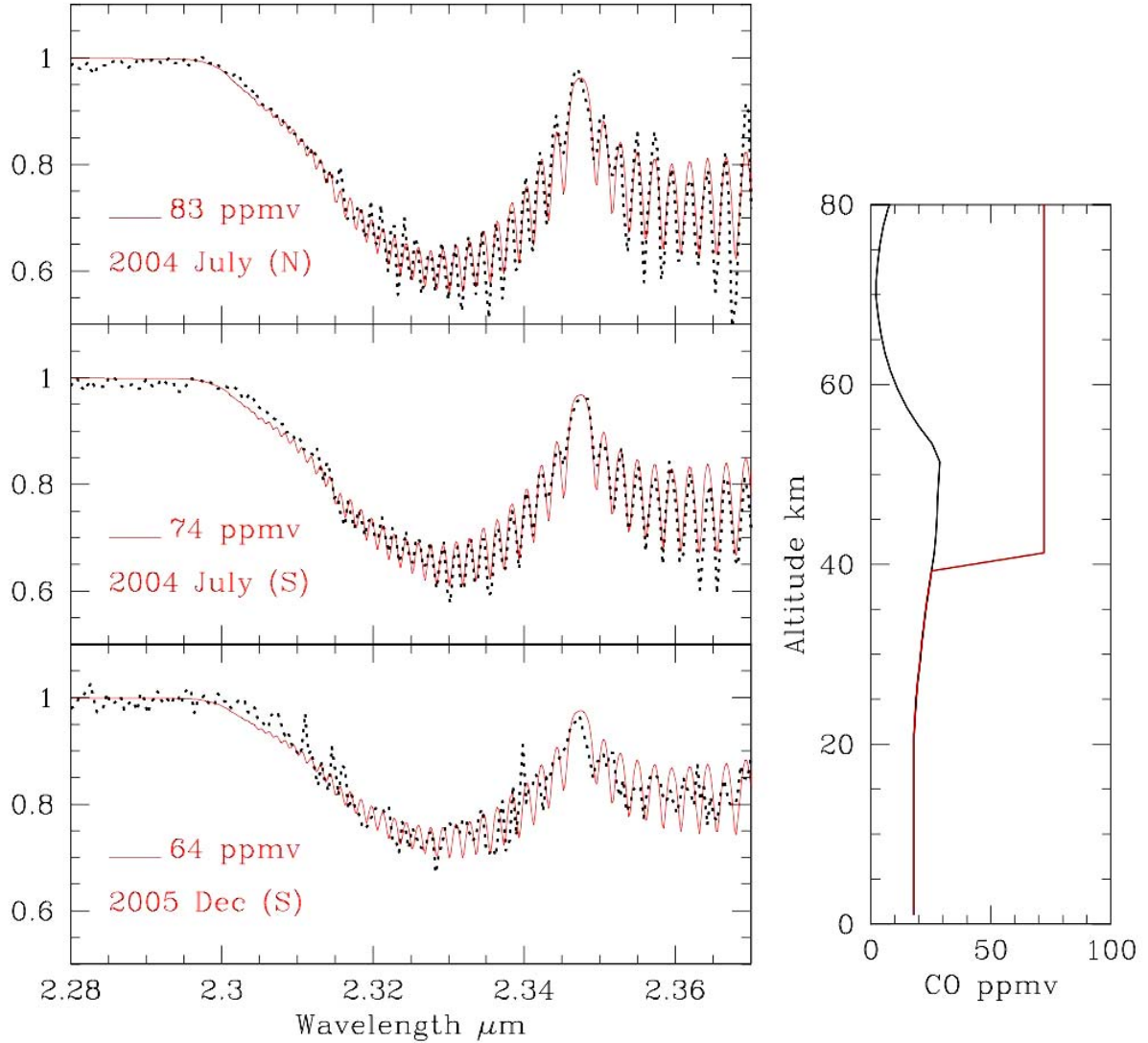

*Fig. 14: Ratio spectra of high to low CO regions (as in Fig. 13 dotted lines) fitted with models in which the CO mixing ratio in the high latitude high CO regions has a fixed value above 40 km and extending to the cloud tops (although we are not very sensitive to CO above about 50 km) as shown in the red line at right. The black line shows the nominal profile that is assumed to apply in the equatorial regions.*

## 5.2 Vertical Distribution

The results in Section 4.3 show that the excess CO in the high latitude regions is centered at an altitude of about 45 km, not around 35 km as most previous studies have assumed. Since our sensitivity to CO is substantially less at 45 km than 35 km that means that a much higher CO mixing ratio is needed to fit the observed spectrum than is the case for CO at lower altitudes. It also means that while the distribution of CO in our maps is accurate, the mixing ratio scale is not representative of the actual situation, because it is based on the assumption of a fixed vertical profile of CO which is scaled equally at all altitudes. This is also true of the results of Marcq et al. (2005, 2008) and Tsang et al. (2008, 2009), as well as the primary determinations presented by Marcq et al. (2006).

We have shown that the data for the high latitude regions is well fitted with a model where all the extra CO is added in a single 2 km high layer. This is, of course, unrealistic. However, any model that has the average height of the CO at about 45 km and doesn't place a significant amount of the extra CO at altitudes below about 36 km fits the data reasonably well. We have made a series of models using a constant CO mixing ratio in the high-latitude region above a specific altitude, normally either 38 or 40 km. Because our sensitivity to CO becomes very low for altitudes above 50 km, it makes little difference whether we add this CO over the region from 38 to 50 km, or extend it all the way from 38 km to the cloud tops. Fig. 14 shows the results of fitting models in which we used a constant CO mixing ratio above 40 km, and our nominal CO profile below that altitude. To match the observed depths of the CO absorption in the ratio spectra we need CO mixing ratios of 83, 74 and 64 ppmv for the data of 2004 Jul (N), 2004 Jul (S) and 2005 Dec (S), respectively. The fits are almost as good as those with a single CO layer (for 2004 Jul(N) the reduced $\chi^2$ is 1.54 for this case compared with 1.51 for a model with all the excess CO in the 45 km layer). If we allow the enhanced CO region to begin at 38 km these numbers become 67, 62 and 54 ppmv, but the fit to the CO band shape is slightly poorer. Models with the extra CO continued to lower altitudes produce much poorer fits to the ratio spectra. Interestingly these CO mixing ratios of around 60 ppmv are similar to those measured at the cloud tops (Iwagami et al., 2010; Krasnopolsky, 2010; Irwin et al., 2008).

## 5.3 Hadley Circulation

These results are therefore consistent with the interpretation for the CO latitudinal variation as being due to a Hadley type circulation with the descending branch occurring at 60 ºN/S (Tsang et al., 2008). CO-rich gas from the upper atmosphere (where CO is created by photo-dissociation of $CO_2$) is carried downwards until it reaches the cloud base at around 48 km. Our results indicate that the CO mixing ratio at this altitude in the high latitudes, is similar to that at the cloud tops.

Below the cloud base CO can be chemically destroyed by reactions such as (Krasnopolsky and Pollack, 1994; Taylor 1995):

$CO + SO_3 \rightarrow CO_2 + SO_2$                                                            (1)

$SO_3$ is relatively abundant at altitudes from 35 to 45 km where it is produced by thermal dissociation of $H_2SO_4$. According to Krasnopolsky and Pollack (1994) this reaction peaks at an altitude of 37 km. So this reaction provides a natural explanation for the CO vertical distribution we are measuring. Another reaction that can also contribute to removal of CO is (Krasnopolsky and Pollack, 1994; Taylor 1995):

$3CO + SO_2 \rightarrow 2CO_2 + OCS$                                                (2)

A recent attempt by Marcq et al. (2008) to use VIRTIS-H to measure latitudinal variation in $SO_2$ below the clouds, which might have provided supporting evidence, was hindered by the narrowness of the spectral band at the long wavelength edge of the instrumental K-band. Consequently only a low accuracy in their measurement was possible, and they were unable to identify a latitudinal trend. Overall they reported 130 ± 50 ppmv at 35 km, between 55 ºN and 50 ºS (Marcq et al., 2008), in line with an earlier ground-based determination (Bézard et al., 1993) of 130 ± 40 ppmv at 35 to 45 km (from an average of four 5 arc-sec diameter aperture measurements centered at 15 ºS, 5 ºN, 25 ºN and 30 ºN), and probe data.

## 5.4 Relationship Between CO and OCS

Previous studies by Marcq et al. (2005, 2006, 2008) have reported a 1:1 anti-correlation of CO at 36 km with OCS at 33 km altitude with latitude. On this basis models that utilize an "unidentified reaction" to convert CO to OCS at these altitudes, most notably Yung et al. (2009), have been developed. However, at the altitudes where we detect the extra CO, the concentration of OCS is significantly less – Bézard et al. (1990) report 0.25 ppmv at 50 km. The extra CO detected using the 2.3 μm window, therefore, can not be directly related to the OCS variations detected using the same window.

Page 25

The prediction of Yung et al.'s (2009) 'Model F' is plotted against our data in Fig. 11, and is not a good match, particularly at high latitudes. Model F, in general, underestimates the CO concentration, but also doesn't predict the polar collar overturning. Here, however, it needs to be remembered that the extra CO we detect is not actually at 35 km, but rather in the range 38 to 50 km. Deeper in the atmosphere CO and OCS might be related, but our 2.3 μm window measurements aren't revealing any CO variations that aren't dwarfed by those higher in the atmosphere. We note that Yung et al.'s (2009) 'Standard Model' (not shown) has a much flatter CO distribution with latitude (2 ppmv equator to pole), but that Model F is the result of tweaking the model parameters to better fit the data of Marcq et al. (2006). As Marcq et al.'s (2006) data comes from similar measurements to our own, it seems that Yung et al.'s (2009) model will, at least, need to be revised.

## 5.5 Dawn-Dusk Asymmetry

One zonal trend identified in the literature is a Dusk-Dawn asymmetry (Tsang et al., 2008). Tsang et al.'s CO maps of the southern hemisphere nightside showed, especially at the polar collars, an increase in CO concentration from West to East (dawn to dusk). In an attempt to confirm that observation, we have merged all of our observations and plotted them on a Local Time-Latitude grid (Fig. 15).

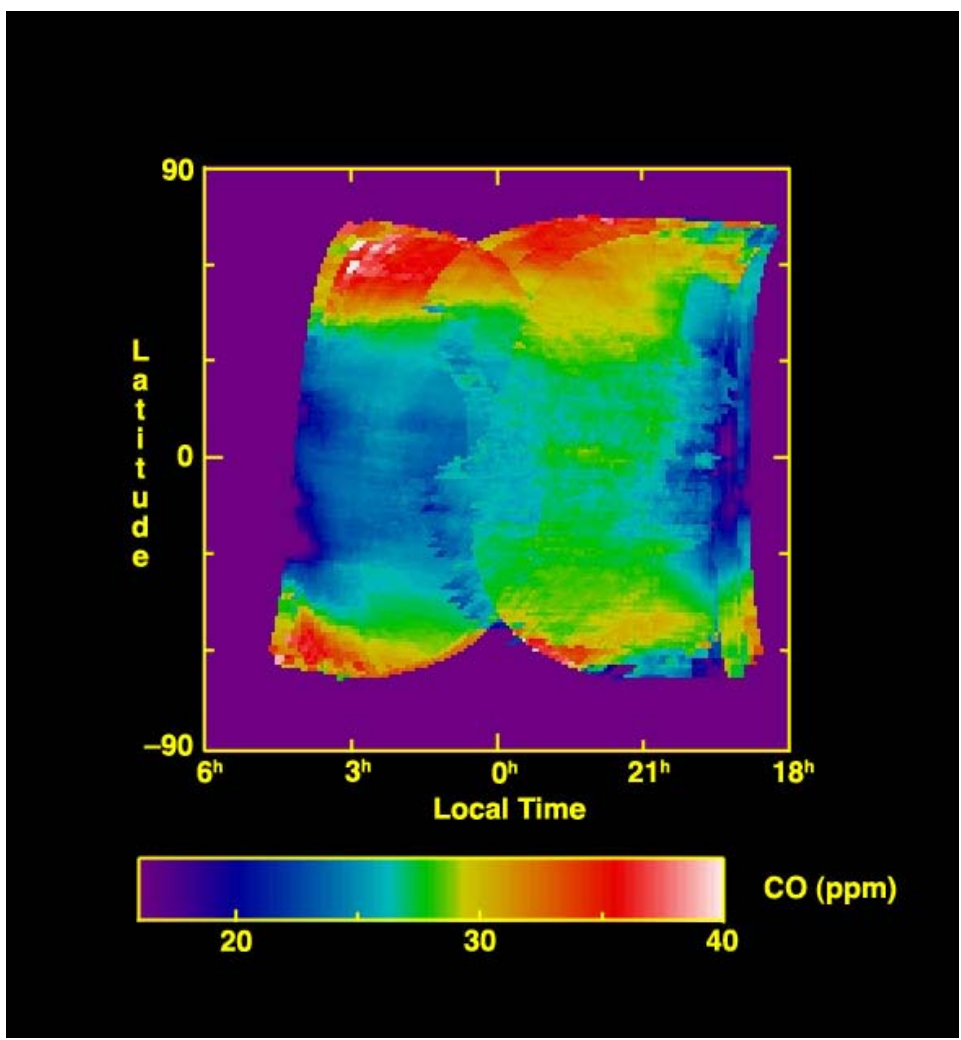

*Fig. 15: CO abundance in ppmv plotted on a Local Time-Latitude grid (excluded data as per Fig. 11).*

The first thing to note is that the edges of the individual observing runs are easily discernable in the merged Fig. 15, which indicates local time is not the dominant cause of variation. However, we do not see the same trend as Tsang et al. (2008). Instead CO concentration at the polar collars appears to fall away from West to East. Though it seems to build up at equatorial latitudes until about 21 LT, we must be cautious here, as only the earliest set of observations (July 2004) covers just after dawn. This set of observations has very little Venusian cloud cover, and therefore this set of CO determinations is slightly low compared to other observations. Additionally, the Early July 2007 run makes a significant contribution to the enhanced CO contribution between 21 LT and midnight, and this is unlikely to be a persistent real feature. However, we still see a fairly substantial drop-off in CO from about 19.30 LT to dusk, which, although not universal in the individual maps of Fig. 9, is a feature frequently discernable in the polar collars. As mentioned in Section 4.1, this detail is also present in one of Tsang et al.'s (2008) maps, and seems likely to be real and frequently, if not continuously, present.

## 5.6 Correspondence to Topography

In Fig. 11 a local maximum in CO concentration, centered just south of the equator is apparent. Similar features are present in observations reported by others, once centered just north of the equator (Marcq et al., 2005, 2006), and twice just south of the equator (Marcq et al., 2008, Tsang et al., 2009) where it was noted that this feature corresponded, to the latitude of Aphrodite Terra (Marcq et al., 2008).

At the altitudes our CO measurements are sensitive to there exists of a stable layer that lends itself to the propagation of gravity waves (Peralta et al., 2008; Gierasch, 1987). Above the orography deficient southern mid- to polar-latitudes, zonally propagating gravity waves have been identified at the top of this layer (~47 km) with horizontal wavelengths in the range 60 to 150 km (Peralta et al., 2008). Modeling studies (Yamamoto and Takahashi, 2009) also predict orographically forced planetary-scale stationary waves corresponding to the latitude of Ishtar Terra, with smaller scale waves corresponding to Aphrodite Terra, at altitudes below 65 km and 50 km respectively. If such planetary-scale structures are present in the lower atmosphere we might expect them to influence the pattern of CO concentration in an observable way.

The expansive northern Ishtar Terra (which includes the extremely tall Maxwell Montes), and the (much lower) southern Lada Terra are situated at latitudes corresponding approximately to the polar collars. Whereas the similarly expansive Aphrodite Terra is situated, for the most part, just south of the equator but offset ~150 ºE from the other two landforms. In Fig. 11, these latitudes correspond to the greatest variation in CO concentration. Such variations have been seen previously at the northern and southern polar collars and equatorial latitudes (Tsang et al., 2009).

The most westward data (relative to the prime meridian) comes from the Early July 2007 run, this part of Venus is not often observed because the planet is further away than at conjunction, and less of the night-side is visible. This fact, taken alone, might lead one to suspect longitudinal topography as the prime candidate for the unusual features. Contrarily, however, the areas of overlapping longitude in the two July 2007 runs are not a good match. Neither is the December 2005 map, which overlaps parts of both July 2007 maps, a good match for either.

The July 2004 run is entirely responsible for data in the eastern hemisphere. The easterly most sections of the data have the polar caps overlaying the western halves of Ishtar Terra and Lada Terra. The highest concentrations of CO we see are the portions of the map closest to (~30 º east of) Maxwell Mons (which is at 0 º longitude). However, similarly high CO concentrations are found in the southern polar collar of the December 2005 data, which, geographically, is nowhere near any significant orography. Similarly the CO maximum in the northern part of the Early July 2007 map corresponds to Atalanta Planitia. We, therefore, do not find any evidence for features in CO corresponding to topographical features at polar latitudes either.

That much variation occurs at the latitudes of the large Terra is likely to be a coincidence resulting from these mountainous regions corresponding to significant atmospheric latitudes. The zonal variations in CO must therefore have another origin.

# 6 Conclusions

Imaging spectroscopy of the nightside of Venus in the 2.0 to 2.4 μm region taken with IRIS2 on the Anglo-Australian Telescope has been used to study the distribution of CO in the Venus atmosphere. We used data spread over four observing runs. We use a CO index defined by the ratio of the intensity in the 2.32 to 2.33 μm region in the CO band to that in the 2.28 to 2.29 μm region outside the band as a measure of CO concentration. Models calculated with the radiative transfer model VSTAR are used to relate the CO index to the CO volume mixing ratio. A number of tests have been carried out to show that the CO index is a reliable measure of CO and is not strongly affected by other factors such as cloud opacity and viewing angle. We used a low intensity cut-off in CLO predominantly to negate the difficulties in crescent subtraction, though this also represents a slight improvement over the technique of Tsang et al. (2009), since the affects of significant cloud cover are removed.

Our maps of the CO distribution confirm the results of previous studies in showing an enhancement of CO at latitudes of 60 ºN/S. However, we see significant variability in the details of the distribution pattern. Where observations were made over several nights we see little indication of night-to-night variability in the distribution, but substantial changes are seen in the pattern over timescales of 20 days or longer.

We have used the ratio of the spectra in regions of high CO concentration at high latitudes with low CO regions near the equator to determine the typical altitude of the excess CO responsible for the enhancements. Modelling with VSTAR shows these spectra are quite sensitive to the altitude of the additional CO absorption. The spectra indicate that the additional CO must be at altitudes of around 45 km, and not at the 35 km altitude as has been assumed in previous studies. Because we are less sensitive to CO at 45 km than at 35 km the amount of extra CO needed in these regions (in ppmv) is higher than has been determined previously. If we assume that the CO in the high latitude regions has a constant mixing ratio at altitudes above 40 km, a mixing ratio in the range 54 to 83 ppmv is needed (compared with 27.5 ppmv at 45 km in our nominal profile which presumably applies to the equatorial regions). This is similar to CO mixing ratios measured at the cloud tops by CO absorption measurements on the dayside.

We suggest that these high CO regions are the result of the downward branch of a Hadley type circulation bringing down CO rich gas from higher in the atmosphere at these latitudes. Once it reaches the regions below the cloud base, CO can be destroyed by chemical reactions with $SO_3$, which is formed here by thermal dissociation of $H_2SO_4$, and with $SO_2$. This leads to a natural explanation for the vertical distribution of CO inferred by our observations with most of the excess CO being above ~40 km.

# 7 Acknowledgements

The authors wish to thank Sarah Chamberlain for her role in the data acquisition. We thank the AAO staff for their assistance with the observations and the AAT time assignment panels for allocation of AAT time. We would also like to thank the Astronomy and Astrophysics group at Macquarie University for their kind donation of space and resources during the early stages of this project.